\documentclass[useAMS,usenatbib,usegraphicx]{mn2e}

\def\spose#1{\hbox to 0pt{#1\hss}}
\def\approxlt{\mathrel{\spose{\lower 3pt\hbox{$\sim$}}
        \raise 2.0pt\hbox{$$<$$}}}
\def\approxgt{\mathrel{\spose{\lower 3pt\hbox{$\sim$}}
        \raise 2.0pt\hbox{$>$}}}
\newcommand{\gta}{\mathrel{\spose{\lower 3pt\hbox{$\mathchar"218$}}
      \raise 2.0pt\hbox{$\mathchar"13E$}}}

\title[Rapid variability in blazars]{Implications of very rapid TeV variability in blazars}

\author[M. C. Begelman, A. C. Fabian \& M.J. Rees] {Mitchell
  C. Begelman $^{1, 2}$\thanks{E-mail: mitch@jila.colorado.edu (MB);
    acf@ast.cam.ac.uk (AF); mjr@ast.cam.ac.uk (MR)}, Andrew
  C. Fabian$^{3\star}$, and Martin J. Rees$^{3}$\footnotemark[1]\\
  $^{1}$ JILA, University of Colorado, Boulder, CO 80309-0440, USA \\
  $^{2}$ Department of Astrophysical and Planetary Sciences,
  University of Colorado at Boulder, CO, USA \\ $^{3}$ Institute of Astronomy,
  Madingley Road, Cambridge CB3 0HA }

\begin{document}
\maketitle

\begin{abstract}
  We discuss the implications of rapid (few-minute) variability in the
  TeV flux of blazars, which has been observed recently with the HESS
  and MAGIC telescopes. The variability timescales seen in PKS
  2155--304 and Mrk 501 are much shorter than inferred light-crossing
  times at the black hole horizon, suggesting that the variability involves enhanced emission in a small region within an outflowing jet. The enhancement could be triggered by dissipation in part of the black hole's magnetosphere at the base of the outflow, or else by instabilities in the jet itself. By considering the energetics of the observed flares, along
  with the requirement that TeV photons escape without producing
  pairs, we deduce that the bulk Lorentz factors in the jets must be
  $\ga 50$.  The distance of the emission region from the central
  black hole is less well-constrained.  We discuss possible consequences for
  multi-wavelength observations.
  
\end{abstract}


\section{Introduction}

The discovery of variable GeV emission from blazars, by Compton
Gamma-Ray Observatory's EGRET instrument (Hartman et al 1992, 2001),
opened up a new way to study the speed, composition, and energetics of
relativistic jets. The multi-day variability timescales measured by
EGRET fit comfortably within the prevailing paradigm that variablity
would be imprinted on the scale of the central black hole's horizon,
and therefore $t_{\rm var} \sim r_g /c = GM/ c^3 = 1.4 m_9$ hr, where
$r_g$ is the gravitational radius and $m_9 = M/ 10^9 M_\odot$ is the
black hole mass in fiducial units.  Disturbances created near the
black hole could travel outward with a high Lorentz factor $\Gamma$ (a
combination of bulk and pattern speed), before radiating energy at a
distance $\ga \Gamma^2 r_g$.  According to this picture, gamma-rays
would be produced at $\sim (10^2-10^4) r_g $, i.e., in a region
approaching the scales where most of the radio emission is produced.
The required Lorentz factors, $\Gamma \la 10$, are also consistent
with the values inferred from radio measurements of superluminal
motion and brightness temperatures. Moreover, placing the GeV emission
region this far from the black hole greatly alleviates the problem of
how the photons escape without producing pairs on the soft photon
background.

This view has now been challenged by the results of TeV observations,
which indicate strong variability in at least two blazars
(PKS 2155--304: Aharonian et al. 2007; and Mrk 501: Albert et
al. 2007a), on timescales as short as a few minutes.  Given the
inferred black hole masses of $\sim 10^9 M_\odot$, these timescales
are one to two orders of magnitude shorter than the shortest
timescales expected.  Although ultrarelativistic motion toward the
observer can preserve a short variability timescale even when the
emission region is far from the black hole, it cannot shorten the
variability timescale imprinted by a source that is stationary in the
observer's frame, without implausible fine-tuning.  Therefore, the TeV
results indicate that the observed variability is imprinted either by
a small fraction of the black hole's horizon, or by small-scale
fluctuations intrinsic to the jet itself.

In this paper we adopt the view that, irrespective of how the variability is triggered, it is the jet itself that is
producing the TeV flares, and study the implications of
short-timescale variability.  In section 2 we present basic scaling
relations that govern the size and energetics of the flaring regions,
and we place these constraints in the context of possible emission
mechanisms in section 3.  We show that the large apparent luminosity
of the flares, and their short timescales, constrain the energy
content of the emitting regions. These constraints, and the requirements that the gamma-rays escape (section 4), indicate that the flaring regions have bulk Lorentz factors $\ga 50$, and most likely produce TeV gamma-rays via Comptonization of external radiation.
 
\section{Size and energetics of flaring regions}

We assume that each flaring region, which has a size $\ell'$ in the
jet comoving frame, is causally connected during the flare (of
comoving duration $\Delta t'$), implying $\ell' < c \Delta t'$.  In
the lab frame the timescale is dilated to $\Gamma \Delta t'$, but from
the observer's point of view the duration of the flare is compressed
by a factor $(1- \beta \cos\theta)^{-1}$, where $\beta = v/c$ is the dimensionless speed and $\theta$ is the angle of motion with respect to the line of sight.  In the limit $\beta \approx 1$ and $\theta \la \Gamma^{-1}$, the compression factor approaches $2\Gamma^2$.  There is a strong observational selection effect that favors this limit for the fastest and most luminous flares, and we will assume that it holds in the following analysis. Thus, an observed variability timescale
$t_{\rm var}$ implies
\begin{equation}
\label{ellprime}
\ell' < t_{\rm var} c \Gamma .
\end{equation}

In the lab frame, the total energy content of the flaring region is
related to the comoving energy density $\varepsilon'$ by $E \sim
\varepsilon' \ell'^3 \Gamma$.  If a fraction $f$ of this energy is
radiated during the flare, the observed power is $P_{\rm obs} \sim f
E/ t_{\rm var}$ and the observer deduces an equivalent {\it isotropic}
luminosity (i.e., luminosity inferred from the observed flux without
accounting for intrinsic anisotropy or beaming effects) of
\begin{equation}
\label{Liso}
L_{\rm iso} \sim 4 P_{\rm obs} \Gamma^2 < 4 f \varepsilon' t_{\rm var}^2 c^3 \Gamma^6 .
\end{equation}
The factor of $\Gamma^2$ in the first relation comes from the fact
that the power is beamed in the direction of the observer.  Measured
values of $L_{\rm iso}$ can be quite large; for example, $L_{\rm iso}
> 10^{46}$ erg s$^{-1}$ for the bright TeV flares observed in
PKS 2155--304 (Aharonian et al. 2007).  We will therefore normalize
$L_{\rm iso}$ to $10^{46}$ erg s$^{-1}$. However, the observed TeV
spectra are quite steep (typical photon spectral indices $\sim
2.5-3.5$), while EGRET spectra suggest that blazar spectra often peak
in the 10--100 GeV band.  Therefore, if the GeV emission flares as
rapidly as the TeV emission, $L_{\rm iso}$ could well be larger than
our fiducial value.
 
The shortest variability timescales that have been measured to date in
PKS 2155--304 (Aharonian et al. 2007) and Mrk 501 (Albert et al. 2007a)
are 3--5 minutes, so we set $t_{\rm var} = 300 t_5$ s.  We then obtain
a lower limit to the internal (comoving) energy density in the flaring
region,
\begin{equation}
\label{epsprime}
\varepsilon' > 10^9 f^{-1} L_{46} t_5^{-2} \Gamma^{-6} {\rm erg \ cm^{-3} } \sim f^{-1} U_r' , 
\end{equation}
where $U_r'$ is the comoving radiation energy density associated with
the flare. Note that $U_r'$ is smaller than the internal energy
density.  

We can assess whether the inferred internal energy density is
reasonable by comparing it to the energy density associated with the
jet flow,
\begin{equation}
\label{lj}
\varepsilon_j' \sim {L_j \over c\Gamma^2 r^2 \Omega} \ga \varepsilon',
\end{equation}
where $L_j$ is the jet power, $r$ is the distance from the black hole
and $\Omega$ is the solid angle subtended by the jet at $r$.  Given the large values of $\Gamma$ that we will deduce for the flaring regions, we anticipate that jet opening angles may have to be be much larger than $\sim \Gamma^{-1}$, in order to explain the statistics of observable sources.  We therefore normalize    
$\Omega$ to 0.1 sr. Setting $x\equiv r/r_g$, we obtain
\begin{equation}  
\label{gammamin}
\Gamma > 1.4 \left( {L_{\rm iso}\over L_j}\right)^{1/4}\left( {\Omega_{0.1}\over f}\right)^{1/4}\left( {x_{\rm fl} m_9\over t_5}\right)^{1/2} ,
\end{equation}
where $x_{\rm fl}$ denotes the location of the flare.  This relation places a significant constraint on $\Gamma$ only if the rapid flares are
produced at radii $\ga 10^3 - 10^4 r_g$.

If the flaring regions are indeed moving outward with the bulk
Lorentz factor of the flow, then 
\begin{equation}
\label{flarerad}
x_{\rm fl} > {c^3 t_{\rm var} \over GM} \Gamma^2 = 6 \times 10^{-2} m_9^{-1} t_5 \Gamma^2 .
\end{equation}
For the values of $\Gamma$ estimated below ($\Gamma \ga 50$), this
would imply $x_{\rm fl}\ga 100$.  However, it is also
possible that the flaring regions are patterns that are fixed relative to the lab
frame (e.g., associated with some external disturbance such as the
funnel of the accretion flow), in which case the emitting gas could be located
closer to the black hole. If the flares are moving outward, we can combine equations (\ref{gammamin}) and
(\ref{flarerad}) to place a lower limit on the jet power, which is
independent of the flare timescale:
\begin{equation}
\label{ljmin}
L_j > 1.4 L_{\rm iso} \left( {\Omega_{0.1}\over f}\right) .
\end{equation}

\section{Radiation mechanisms}

The double humped SEDs of blazars are generally attributed to the superposition of a synchrotron spectrum --- peaking in
the IR--optical or UV--X-ray for ``low-peaked blazars" (LBLs) and ``high-peaked blazars" (HBLs), respectively --- and an inverse Compton spectrum (peaking at gamma-ray energies) produced by the same population of electrons.  Both PKS 2155--304 and Mrk 501 are HBLs, as indeed are all but one of the blazars detected at TeV energies, to date.  

The seed photons for Comptonization are provided primarily by either the synchrotron photons themselves (the synchrotron self-Compton [SSC] mechanism: Maraschi, Ghisellini \& Celotti 1992; Bloom \& Marscher 1996) or a radiation field impinging on the jet from outside (the External Radiation Compton [ERC] mechanism: Begelman \& Sikora 1987; Melia \& K\"onigl 1989; Dermer \& Schlickeiser 1994; Sikora, Begelman \& Rees 1994).   Since the flare radiation density in the comoving frame, $U_r'$ (eq.~[\ref{epsprime}]), includes both the synchrotron and Compton components, the dominance of the Compton (gamma-ray) hump strongly favors the ERC mechanism over the SSC mechanism.  On the other hand, one cannot make such a strong statement if the synchrotron peak dominates; in this case, either mechanism is viable. The SEDs of the two highly variable TeV sources do not exhibit a trend: in Mrk 501 the synchrotron peak appears to dominate (Albert et al. 2007a), while the Compton component dominates the SED of PKS 2155--304 in data presented by Foschini et al. (2007; but not in the [non-simultaneous] data quoted by  Ghisellini et al. 1998).   

The fact that the spectra are quite flat (or even inverted) longward of each peak, and quite steep shortward, suggests that most of the energy in the accelerated electrons is contained in particles with high random Lorentz factors, radiating close to the peak.  This view contrasts with earlier assertions that the peak is associated with cooling of the electrons (Sikora et al. 1994; Ghisellini et al. 1998), and could mean that the acceleration mechanism is ``particle-starved", in the sense that the Poynting flux  exceeds the kinetic energy flux in the flaring region (Sikora et al. 2005).  If particle acceleration is efficient in these flares (i.e., if $f$ is not too small and most electrons are
accelerated), then we may assume that the intrinsic energy density is
primarily magnetic, $\varepsilon \sim B'^2/ 8\pi$, with  
\begin{equation}
\label{Bprime}
B' > 2\times 10^5 f^{-1/2} L_{46}^{1/2} t_5^{-1} \Gamma^{-3} {\rm G} .  
\end{equation} 
According to this emission model, the intrinsic synchrotron emissivity
of PKS 2155--304 peaks at $\nu_{\rm syn} \sim 10^{16} \nu_{16}/\Gamma $ Hz, which
requires the random Lorentz factors of electrons contributing to the
peak to satisfy
\begin{equation}
\label{gampeak}
\gamma_{\rm peak} < 250 \nu_{16}^{1/2}f^{1/4} L_{46}^{-1/4} t_5^{1/2} \Gamma  .  
\end{equation} 
If the intrinsic gamma-ray spectrum, peaking at $\sim 10^{24}/ \Gamma$
Hz, is produced by Comptonization of the synchrotron spectrum, then
$\gamma_{\rm peak} \sim 10^4$ and $\Gamma \ga 50$.  In this
picture, scattering in the Klein-Nishina regime could contribute to the steepness of the TeV
spectrum. 

Although large values of $\Gamma$ may be needed to produce rapidly
fluctuating gamma-rays, they can also inhibit efficient synchrotron cooling of the
flare plasma. We have implicitly assumed that all of the dissipated
energy is radiated away during the flare.  This implies that the
cooling timescale in the comoving frame is shorter than $\Gamma t_{\rm
  var}$ --- if this were not the case then our energetic requirements
would have to increase.  If cooling is dominated by synchrotron
losses, then a sufficient condition to ensure efficient cooling is
\begin{equation}
\label{coolcondition1}
\gamma > 10^{-4} f L_{46}^{-1} t_5 \Gamma^5 .
\end{equation}
To ensure that all electrons with $\gamma > \gamma_{\rm peak}$ are able to cool, we require
\begin{equation}
\label{coolcondition2}
\Gamma < 40 \nu_{16}^{1/8}f^{-3/16} L_{46}^{3/16} t_5^{-1/8}  .  
\end{equation} 
Thus, an SSC model for rapid flares from PKS 2155--304 can barely
satisfy the condition for efficient cooling above $\gamma_{\rm peak}$,
given our deduction that $\Gamma \ga 50$; this constraint
will become tighter once we consider the opacity due to pair
production.

Constraints on radiative efficiency are relaxed considerably if the gamma rays are produced by the ERC mechanism.  In order for ERC to dominate over SSC, the ambient radiation energy density must exceed the synchrotron energy density as measured in the comoving frame.  The external radiation density in the lab frame must then satisfy
\begin{equation}
\label{Uext}
U_{\rm ext}> 10^9 L_{46} t_5^{-2} \Gamma^{-8} {\rm erg \ cm^{-3} } , 
\end{equation}
if the intensity at $x_{\rm fl}$ is approximately isotropic, corresponding to a luminosity $L_{\rm ext} \sim 2\times 10^{35} (\Gamma/50)^{-8} m_9^2 x_{\rm fl}^2$ erg s$^{-1}$. Henceforth we normalize $\Gamma$ to 50 because both energetic and transparency constraints will demand such values. If the external radiation illuminates the flaring region from behind, subtending a small solid angle $\Omega_{\rm ext}$ (with $\Gamma^{-2}\ll \Omega_{\rm ext}/2\pi \ll 1 $) in the lab frame, then the required energy density and luminosity are increased by a factor $(\Omega_{\rm ext}/2\pi)^{-2}$.  For radiation emitted at $r_{\rm ext} \ll r_{\rm fl}$ ---  from the central region of an accretion disk, for example --- this factor is $\sim (r_{\rm ext}/ r_{\rm ext})^4$, much steeper than the expected variation of emissivity with radiation in most accretion scenarios.  Therefore, it seems most likely that the external radiation responsible for Comptonization is emitted at radii comparable to $x_{\rm fl}$ (Dermer, Schlickeiser \& Mastichiadis 1992; Dermer \& Schlickeiser 1993).

If the ($\sim$ isotropic) external radiation field peaks at a frequency $\nu_{\rm ext}$, then the condition for producing the gamma-ray peak at $\nu_\gamma \sim 10^{24}$ Hz is 
\begin{equation}
\label{ERC}
{\nu_\gamma \over \nu_{\rm ext} }\sim \Gamma^2 \gamma_{\rm peak}^2 .
\end{equation} 
Combining this condition with eq.~(\ref{gampeak}), we obtain
\begin{equation}
\label{nuext}
\nu_{\rm ext} > 2 \times 10^{12} \nu_{16}^{-1} f^{-1/2} L_{46}^{1/2} t_5^{-1} \left( {\Gamma\over 50}\right)^{-4} \ {\rm Hz } .
\end{equation}
Thus, an ERC model for rapidly varying TeV flares would require an external radiation source in the submillimeter band, if the external radiation is diffuse, and a factor $\sim (\Omega_{\rm ext} /2\pi)^{-1}$ higher frequency if it comes from behind.

In contrast to the SSC mechanism, the cooling of relativistic
electrons in the ERC model becomes more efficient with increasing
$\Gamma$, with the least efficient cooling occurring when the two
mechanisms are comparable.  For electrons near $\gamma_{\rm peak}$ the
ratio of cooling time to variability time is sensitive to $\Gamma$,
varying $\propto \Gamma^4$ when SSC dominates and $\propto \Gamma^{-4}$
when external radiation controls energy loss.
 
\section{Escape of radiation} 

The most stringent conditions on our flare model are set by the
requirement that the gamma-rays escape without being absorbed in
$\gamma-\gamma$ pair production.  In blazar models there are two
possible targets for pair production: the synchrotron radiation
intrinsic to the jet and the external ambient radiation.  In the case
of interactions with jet radiation, optimal conditions for escape
occur if a TeV photon encounters radiation only within the flaring
region in which it was produced.  If the TeV photon has to pass
through other radiating regions, this will decrease its escape
probability; but given the amplitude of observed flares it is
plausible that the escape constraint is set locally.\footnote[1]{Note, however, that in the Sikora
et al. (1994) ERC model for 3C 279, the pair
production constraint due to internal synchrotron radiation is much
less severe than that due to external diffuse radiation.} In
considering pair production on ambient radiation, on the other hand,
one must integrate over a path length of order $r$.

To estimate the pair production opacity internal to the flare region,
we use the estimate of radiation density $U'_r$ from
eq.~(\ref{epsprime}).  We assume that this energy density is dominated
by synchrotron radiation peaking at a frequency $10^{16}
\nu_{16}/\Gamma$ Hz, which therefore has a number density
\begin{equation}
\label{nnupeak}
\nu_{\rm peak} n'_{\nu}(\nu_{\rm peak}) \sim 6 \times 10^{19} \nu_{16}^{-1} L_{46} t_5^{-2} \Gamma^{-5} \ {\rm ph \ s}^{-1}.
\end{equation}
The cross section for pair production peaks at
$\sigma_{\gamma\gamma}\sim \sigma_T/5$ close to threshold, where
$\sigma_T$ is the Thomson cross section, and declines at higher
energies.  Blazar spectra tend to be quite flat longward of the
synchrotron peak, with an energy spectral index $\alpha \sim 0.3-0.5$
(where the flux is $F_\nu \propto \nu^{-\alpha}$).  Shortward of the
peak they decline somewhat more steeply than $\nu^{-1}$.  Under these
circumstances, the most probable pair-producing reactions are those
close to threshold.  The likely targets for photons
with energy 1 TeV$/\Gamma$ (i.e., the photons with observed energy
$\sim 1$ TeV) have frequencies $\nu_{\rm target} \sim 4 \times 10^{13}
\Gamma$ Hz.  We therefore need to correct $\nu_{\rm peak}
n'_{\nu}(\nu_{\rm peak})$ by a factor $(\nu_{\rm peak}/ \nu_{\rm
  target})^\alpha$ to account for the ratio of target photons to
photons near the spectral peak.  Anticipating that the minimum $\Gamma
$ for PKS 2155--304 will be large enough that $ \nu_{\rm target} >
\nu_{\rm peak}$, we adopt $\alpha = 1$. The correction
factor is then $\sim 160 \nu_{16}\Gamma^{-2}$.  Multiplying by a path
length $\ell' = c t_{\rm var} \Gamma$ and the threshold cross section,
we obtain the optical depth to pair production,
\begin{equation}
\label{taugamma}
\tau_{\gamma\gamma} \sim 2\times 10^{10} L_{46} t_5^{-1} \Gamma^{-6}.
\end{equation}
The condition $\tau_{\gamma\gamma} \la 1$ then implies
\begin{equation}
\label{taugamma2}
\Gamma > 50 L_{46}^{1/6} t_5^{-1/6} 
\end{equation}
(Celotti et al. 1998). Thus, the escape of TeV photons from the site of a rapid flare requires $\Gamma \ga 50$, provided that the observed synchrotron emission comes from the same site. This result is independent of the distance of the flaring region from the black hole.

In the presence of diffuse external radiation, the threshold target frequency for pair production by a 1 TeV photon is
$\sim 6\times 10^{13}$ Hz, or about an order of magnitude higher than $\nu_{\rm ext}$ given by eq.~(\ref{nuext}), if $\Gamma \sim 50$.
Using eq.~(\ref{Uext}) to estimate the external radiation density, and conservatively assuming a flat spectral index ($\alpha=0$), we
estimate a target photon number density $\nu_{\rm ext} n_{\nu}(\nu_{\rm ext}) \sim 3 \times 10^8 L_{46} t_5^{-2} (\Gamma /
50)^{-8}$ ph cm$^{-3}$. To obtain the pair production optical depth we multiply by the threshold cross section and a path length $\sim r =
1.5 \times 10^{14} m_9 x$:
\begin{equation}
\label{taugammaext}
\tau_{\gamma\gamma, \ ext} \sim 6\times 10^{-3} L_{46} t_5^{-2}\left( {\Gamma\over 50}\right)^{-8} m_9 x_{\rm fl}.
\end{equation}
This relation implies that the rapid TeV variability could also be produced by Comptonization of diffuse external radiation, but only out to a radius of several hundred $r_g$ for $\Gamma \sim 50$.  Higher values of $\Gamma$ would allow the flare to occur at larger radii.

If the external radiation came from behind the jet, this constraint would not be much changed.  In a typical pair-producing collision, the angle between a TeV photon and the soft target would be small, $\sim (\Omega_{\rm ext}/2\pi)^{1/2}$, but because the incident soft photons must be more energetic by a factor $(\Omega_{\rm ext}/ 2\pi)^{-1}$, the pair production threshold is still about an order of magnitude above $\nu_{\rm ext}$.  The ambient radiation density required by the ERC model is larger by a factor $(\Omega_{\rm ext}/2\pi)^{-2}$, but this increase is compensated for by the decrease in target photon number and collision rate per target photon, each of which scales as  $\Omega_{\rm ext}/2\pi$.  Thus, the optical depth for pair production is not much changed, assuming that the ERC mechanism produces the TeV flares. If the SSC mechanism dominates, so that the external radiation supply is weaker, then the pair production constraint would be correspondingly relaxed.   

These results are consistent with the ERC model of Sikora et al. (1994), which focused on the low-peaked blazar 3C 279.  In
that case, the synchrotron peak was at lower frequencies, $\nu_{16} \sim 0.1$, while the variability timescale was taken to be 1
day ($t_5 \sim 300$) with $\Gamma \sim 5$.  These numbers
give an external radiation peak frequency in the ultraviolet,
$\nu_{\rm ext} \sim 10^{16}$ Hz, and an optical depth for pair
production $\tau_{\gamma\gamma, \ ext} \sim 7\times 10^{-4} m_9 x $.
Thus, flares located at about $r \sim 10^{18}$ cm are marginally
optically thin to pair production.

\section{Discussion}

We have analyzed the requirements for producing rapid (few-minute),
high amplitude TeV variability in relativistic blazar jets. Our model
is based on the two standard models for gamma-ray emission from
blazars, in which the two spectral ``humps'' correspond to synchrotron
radiation and inverse Compton scattering, respectively.  As in the
standard models, we show that the flaring gamma-rays could be produced
either by Comptonizing the synchrotron photons (SSC model) or by
Comptonizing a diffuse background source of submillimeter
radiation (ERC model). Since the flaring regions must be quite small
to satisfy causality constraints, they must have large bulk Lorentz
factors, $\Gamma \ga 50$, in order for the TeV radiation to avoid pair
production against the synchrotron photons.  VLBI measurements of superluminal motion in PKS 2155--304 (Piner \& Edwards 2004) and Mrk 501 (Giroletti et al. 2004) suggest lower values of $\Gamma$ in these objects, but we stress that the regions producing radio emission and gamma-ray flares may have very different properties.

Although both SSC and ERC mechanisms appear capable of explaining the rapid TeV variability, ERC seems more likely to dominate.  Whereas SSC models are tightly constrained (or possibly excluded) between upper limits on $\Gamma$ imposed by radiative efficiency and lower limits imposed by transparency and Comptonization constraints, ERC models merely have to satisfy a lower
limit on $\Gamma$ (since the radiative efficiency of ERC increases with $\Gamma$).  The radiation energy densities required by ERC are
modest, and would be hard to avoid, even in the presence of a radiatively-inefficient accretion flow of the sort likely to be found in BL Lac objects.  Moreover, the dominance of the Compton peak over the synchrotron peak, which characterizes the SED of at least one of the highly variable TeV sources (PKS 2155--304: Foschini et al. 2007), is strong, direct evidence for the ERC process. 

ERC flares could be produced close to the $\gamma-\gamma$ photosphere, along the lines suggested by Blandford \&
Levinson (1995).  Setting $\tau_{\gamma\gamma, \ {\rm ext}} \sim 1$ in eq.~(\ref{taugammaext}) and using eq.~(\ref{Uext}) to estimate the external radiation density, we obtain an external luminosity of $L_{\rm ext} \sim 3 \times 10^{40} L_{46}^{-2}t_5^4 (\Gamma/50)^8$ erg s$^{-1}$.  This does not represent the entire unbeamed luminosity of the blazar, but merely the portion produced at radii $\ga x_{\rm fl}$. If the flare occurs at radii $x_{\rm fl} \sim 10^2-10^3$, as seems likely, then such low luminosities could be fully compatible with standard accretion models.  In particular, the external sub-mm radiation could be nonthermal emission produced by a radiatively inefficient accretion flow, or thermal emission produced by the outer, cool parts of an accretion disc. Alternatively, it could be produced by relativistic electrons in a shear
layer surrounding the jet.  The effective external luminosity would
vary with radius, depending on the geometry of the source and the
run of $\Gamma(r)$.  If these behaviors were known, one could
calculate a relationship between the isotropic luminosity of a flare
and its duration.  The measured PDS $\propto freq.^{-2}$ of PKS
2155--304 (Aharonian et al. 2007) implies $L_{\rm iso} \propto t_{\rm
  var}$.

The intrinsic steepness of the TeV spectrum in the rapidly varying
sources is unlikely to arise from cooling, since the
spectrum appears to be considerably steeper than the hard tail of the
synchrotron hump.  Klein-Nishina effects could contribute to the
steepness, as could ``self-absorption'' due to pair production.
The latter effect can occur when the TeV photons pair-produce against synchrotron
photons shortward of the peak. Ignoring Klein-Nishina effects, the
emissivity for TeV gamma-rays has the same slope as the
high-energy synchrotron photons (say, $j_\nu \propto \nu^{-\alpha}$,
where $\alpha \sim 1$).  But since the absorption coefficient scales
with TeV frequency as $\alpha_\nu \propto \nu^{\alpha}$, higher energy TeV
photons find more targets for pair production.  Therefore, if the
source is self-absorbed, the TeV energy spectral index is $2
\alpha$.  This would tend to decrease the TeV variability relative to
the synchrotron photons, if SSC dominates, and might
cancel the nonlinearity expected from the SSC mechanism.

Our analysis implies some significant differences between LBLs and
HBLs that may explain why only the latter have exhibited rapid
variability at TeV energies (and, with the exception of a detection of
weak emission from BL Lac [Albert et al. 2007b], why only HBLs have been
detected in the TeV band). Since LBLs have $\nu_{16} \sim
10^{-4}-10^{-2}$, equation (\ref{gampeak}) implies that 
$\gamma_{\rm peak}$ would have to be lower by a factor $0.01-0.1$ if these
objects were to produce very rapid flares via the dissipation of
magnetic energy. In this case, gamma-rays could not be produced by
SSC; ERC would have to dominate, with the characteristic frequency of the external radiation 
$\nu_{\rm ext}$ in the UV. Such an external radiation source is
probably not found in BL Lac objects, but might be present in optically violently variable (OVV)
quasars.  In the quasar case, however, both the radial scale of the
external radiation and its luminosity are expected to be quite large,
implying that the pair photosphere will be located at $\ga 10^4 r_g$.
At these radii the jet is likely to be so expanded that the energy
density will be insufficient to create an intense, rapid flare
(cf. eq.~[\ref{gammamin}]); moreover, the conversion of Poynting flux
to kinetic energy (or its loss to radiation) may have largely taken place by this radius, further hampering the
production of flares.  Therefore, we might not expect such a system to
produce very rapid flares with high luminosities.  LBL BL Lac objects may have trouble producing rapid flares even at GeV energies, since they lack a strong source of UV radiation; to produce such flares in these objects would require the slow
dissipation of relatively weak magnetic fields, according to equations
(\ref{Bprime}) and (\ref{gampeak}); thus, we would expect $t_5 \gg 1$.
These arguments also suggest a reason for the relative weakness of TeV
emission, compared to GeV emission, in those LBLs that have been
detected in gamma-rays.  In LBLs, the external Comptonization required to
produce TeV photons probably extends further into the Klein-Nishina regime,
since the photon energy in the electron rest frame is roughly the
geometric mean between the seed photon energy and the TeV final
energy.  This will compound the spectral steepening due to effects
like self-absorption due to pair production, creating a much weaker TeV
signal.
 
Our model does not make strong predictions about the distance from the
black hole at which the flares are produced.  The main
constraint is that the flow must have accelerated to $\Gamma \ga 50$
before reaching the flare zone.  It is well known that relativistic
jets, subject to magnetohydrodynamical or fluid forms of pressure, tend
to accelerate rather slowly.  In a ballistic, ultrarelativistic flow
$\Gamma$ increases roughly linearly with radius, suggesting that the
jet should reach at least $100 r_g$ before producing the observed
flares.  If the jet is collimated, e.g., by magnetic stresses, the
acceleration could be slower still.  However, if the jet is extremely
``particle-starved'' --- as we suggested based on evidence that the
mean random Lorentz factor of electrons in the flare zone could be as
large as $\sim 10^4$ --- then it may undergo a very rapid episode of
acceleration close to the black hole.  By analogy with pulsar winds, magnetocentrifugal stresses can accelerate the wind to
Lorentz factors $\sim \sigma^{1/3}$ close to the light cylinder, where
$\sigma$ is the ratio of energy density to rest mass density and the
light cylinder would generally be expected to lie at a few $r_g$ (Michel 1969; Begelman \& Li 1994).  If
this ratio could be as large as $10^6$ for blazar jets, then these
jets would probably accelerate to $\Gamma \sim 100$ almost
immediately.  The remaining energy density would be just enough for
the electrons to be accelerated to an average Lorentz factor of $\sim
10^4$.

The detection of large-amplitude variability on timescales shorter than $r_g/c$ is important and surprising. We can already infer higher Lorentz factors than have generally been contemplated for blazars, and that the phenomenon is triggered by processes that involve extreme relativistic plasmas, perhaps in a black hole magnetosphere. These possibilities lend added motivation to future observations --- and especially to simultaneous multiband observations that could discriminate among the options.

\section*{Acknowledgments}
MCB acknowledges support from NSF grant AST-0307502 and NASA Astrophysics Theory Program grant NNG06GI06G. We thank the referee, Fulvio Melia, for helpful comments.


\begin{thebibliography}{}
\bibitem[]{} Aharonian F., et al., 2007, ApJ, 664, L71
\bibitem[]{} Albert J., et al., 2007a, ApJ, submitted, astro-ph/0702008
\bibitem[]{} Albert J., et al., 2007b, ApJ, 666, L17
\bibitem[]{} Begelman M.C., Li Z.-Y., 1994, ApJ, 426, 269 
\bibitem[]{} Begelman M.C., Rees M.J., Sikora M., 1994, ApJ, 429, L57
\bibitem[]{} Begelman M.C., Sikora M., 1987, ApJ, 322, 650 
\bibitem[]{} Blandford R.D., Levinson A., 1995, ApJ, 441, 79
\bibitem[]{} Bloom S.D., Marscher A.P., 1996, ApJ, 461, 657
\bibitem[]{} Celotti A., Fabian A.C., Rees M.J., 1998, MNRAS, 293, 239
\bibitem[]{} Dermer C.D., Schlickeiser R., 1993, ApJ, 416, 458
\bibitem[]{} Dermer C.D., Schlickeiser R., 1994, ApJS, 90, 945   
\bibitem[]{} Dermer C.D., Schlickeiser R., Mastichiadis A., 1992, A\&A, 256, L27  
\bibitem[]{} Foschini L., et al, 2007, ApJ, 657, L81
\bibitem[]{} Ghisellini G., Celotti A., Fossati G., Maraschi L., Comastri A., 1998, MNRAS, 301, 451
\bibitem[]{} Giroletti M., et al., 2004, 600, 127
\bibitem[]{} Hartman R.C., et al 1992, ApJ, 385, L1 
\bibitem[]{} Hartman, R.C., 2001, ApJ, 558, 583
\bibitem[]{} Maraschi L., Ghisellini G., Celotti A., 1992, ApJ, 397, L5
\bibitem[]{} Melia F., K\"onigl A., 1989, ApJ, 340, 162 
\bibitem[]{} Michel F.C., 1969, ApJ, 158, 727 
\bibitem[]{} Piner B.G., Edwards P.G., 2004, ApJ, 600, 115
\bibitem[]{} Sikora M., Begelman M.C., Madejski G.M., Lasota J.-P., 2005, ApJ, 625, 72 
\bibitem[]{} Sikora M., Begelman M.C., Rees M.J., 1994, 421, 153

\end{thebibliography}
\end{document}